\documentclass[lettersize,journal]{IEEEtran}
\usepackage{amsmath,amsfonts}
\usepackage{mathtools}
\usepackage[linesnumbered,ruled]{algorithm2e}
\usepackage{array}
\usepackage[caption=false,font=normalsize,labelfont=sf,textfont=sf]{subfig}
\usepackage{textcomp}
\usepackage{amsmath}
\usepackage{stfloats}
\usepackage{url}
\usepackage{verbatim}
\usepackage{graphicx}
\usepackage{cite}
\usepackage{xcolor}
\usepackage{ifdraft}
\usepackage{enumitem}
\usepackage{multirow}
\usepackage{amsthm}
\newtheorem{define}{Definition}

\allowdisplaybreaks
\begin{document}
 % \title{ Feasible Region Assessment under Imperfect Information: A Data-Driven Approach for Coal Mine Integrated Energy System Flexibility}
  % \title{ Learning for Feasible Region Assessment: A Data-driven Approach for Coal Mine Virtual Power Plants with Imperfect Information}
  \title{ Learning for Feasible Region on Coal Mine Virtual Power Plants with Imperfect Information}
  \author{Hongxu Huang~\IEEEmembership{Member,~IEEE}, Ruike Lyu,~\IEEEmembership{Graduate Student Member,~IEEE},Cheng Feng,~\IEEEmembership{Member,~IEEE}, Haiwang Zhong~\IEEEmembership{Senior Member,~IEEE},  H. B. Gooi~\IEEEmembership{Life Fellow,~IEEE}, Bo Li ~\IEEEmembership{Member,~IEEE}, and Rui Liang~\IEEEmembership{Senior Member,~IEEE}\vspace{-0.5cm}}

% % \thanks{This paper was produced by the IEEE Publication Technology Group. They are in Piscataway, NJ.}% <-this % stops a space
% % \thanks{Manuscript received April 19, 2021; revised August 16, 2021.}
\markboth{Journal of \LaTeX\ Class Files,~Vol.~14, No.~8, October~2024}%
{Shell \MakeLowercase{\textit{et al.}}: A Sample Article Using IEEEtran.cls for IEEE Journals}

% (Huang Hongxu and Lahanda Purage Mohasha Isuru Sampath contributed equally to this work.) (Corresponding author: Hung Dinh Nguyen.)

\maketitle

\begin{abstract}
The feasible region assessment (FRA) in industrial virtual power plants (VPPs) is driven by the need to activate large-scale latent industrial loads for demand response, making it essential to aggregate these flexible resources for peak regulation. However, the large number of devices and the need for privacy preservation in coal mines pose challenges to accurately aggregating these resources into a cohesive coal mine VPP. In this paper, we propose an efficient and reliable data-driven approach for FRA in the coal mine VPP that can manage incomplete information. Our data-driven FRA algorithm approximates equipment and FRA parameters based on historical energy dispatch data, effectively addressing the challenges of imperfect information.  Simulation results illustrate that our method approximates the accurate feasible operational boundaries under dynamic and imperfect information conditions.

\end{abstract}


\vspace{-1mm}
\section{Introduction}

\IEEEPARstart{T}{HE} evolution of energy markets has highlighted the potential of Virtual Power Plants (VPPs) for flexible energy management and trading. Unlike distributed energy resources, industrial sectors with huge regular power demands present significant opportunities for large-scale VPP aggregation. As a major energy-intensive industry in China, the coal mines depend heavily on electricity for their production processes, contributing to an annual power demand of 95.1 billion kWh in 2022 \cite{coalconsumption}. Given this insight, it is essential to activate the dormant flexibility resources within coal mine industrial energy systems (CMIESs).

Under this circumstance, for VPPs to effectively participate in energy markets, the accurate model of industrial energy system is crucial to describe the both energy consumption behaviors and the flexibility of diverse industrial energy systems. Considerable strides have been achieved in industrial energy system modeling. To name a few, Ref. \cite{lyu2023lstn} focused on linearizing the complex nonlinear non-convex industrial demand response model for VPP aggregation as a virtual battery model \cite{Tan_VB}.  In \cite{Industrial_VPP1}, a multi-energy industrial park model is proposed to participate in day-ahead energy and reserve markets via VPP aggregation, ensuring all possible deployment requests can be realized. To center on coal mine industries, the associated energy recovery in coal mines are considered in \cite{HUANG} to improve operational economic benefits. Further considering the flexibility in coal transportation, the integration of belt conveyors (BCs) and coal silos for demand response (DR) are modeled in \cite{VES} under the energy-transportation coordinated operation framework. Apart from the economic cost, a multi-objective model of CMIESs dispatching are solved with multitask multi-objective algorithm \cite{MaJun}. For participate in energy trading, Ref. \cite{Blockchain_coalmine} proposed a CMIES model for participating in the integrated energy and carbon trading market, while coal mines are treated as individual entities rather than as an aggregated VPP. However, these mentioned models are designed for coal mine energy dispatching, remaining a notable gap in aggregating the flexibility of CMIESs. 

{\color{black}Flexibility aggregation has become a widely researched topic in the field of VPPs. To effectively participate in the energy market, a VPP needs to estimate a feasible region that defines the boundaries within which these aggregated resources can be managed reliably and efficiently. }Various approaches are proposed to manage the feasible region assessment (FRA), such as convex hull outer approximation \cite{JYP} and  the Minkowski sum based estimation method \cite{MKS}. Although these method are effective in FRA problems, they still rely on perfect information of all participants in the VPP aggregation, which neglects the fact that usually all the participants are not willing to share their privacy information. To tackle the imperfect information in FRA, a privacy-preserving based FRA method is proposed in \cite{Pareek} for peer-to-peer energy trading under uncertain renewable energy generation, while it still relies on the probabilistic density function which is hard to get via the model-driven approaches. {\color{black}However, these methods face significant limitations when applied to the CMIESs due to the issue of imperfect information. In CMIES, the vast number of devices and complex interactions between energy units make it difficult to obtain accurate and complete parameter on all CMIESs. The imperfect information of CMIESs makes it exceptionally challenging to aggregate coal mines into a cohesive VPP capable of reliably participating in energy markets. Thus, how to get the accurate feasible region for CMIESs aggregation is still a remaining chanllenge.}

% Meanwhile, in coal mine industry, each CMIES contains various energy equipment for their coal production. The high number of devices and systems within each mining operation, coupled with incomplete or unknown parameters for many of these energy units, further complicates the accurate estimation of the feasible region. The imperfect information of CMIESs makes it exceptionally challenging to aggregate coal mines into a cohesive VPP capable of reliably participating in energy markets. How to get the accurate feasible region for CMIESs aggregation is still a remaining chanllenge.

In order to overcome the imperfect information in FRA problems, data-driven based method has gained as a promising approach to estimate feasible region of complex systems like the CMIES. By utilizing the histrocial optimal dispatching data, the inverse optimization method \cite{IO_1} is developed to infer underlying parameters and constraints that define optimal operational states, allowing for more accurate and adaptive modeling. Ref. \cite{Tan_DDIO} leveraged the structure of virtual battery model and proposed Newtown's method based inverse optimization algorithm for FRA. In \cite{Ruike}, the data-driven inverse optimization is also developed to solve simplified virtual battery based VPP aggregation among diverse EVs. {\color{black}Although data-driven inverse optimization has made progress, existing methods struggle with the unique complexities of CMIESs, such as the complex model of belt conveyors, raw 1coal mining and transportation networks and processes coordination. These systems require real-time, efficient computation under dynamic conditions, which conventional methods still face challenges in computation efficiency and accuracy. Thus, further advancements are needed to adapt these approaches for the specific demands of large-scale CMIESs.}

% However, while data-driven inverse optimization has made notable progress, existing methods still face challenges in computation efficiency and accuracy, which is not applicable in the complex CMIES. Further data-driven advancements are needed to enhance the approach more viable for real-time applications and large-scale CMIESs.

To address the research gaps identified above, this paper proposes the learning-based FRA method for VPP aggregating the CMIESs with imperfect information. The main contributions of this work are summarized as:
\begin{enumerate}
\item To reduce the computational burden of FRA, an inverse optimization model is developed based on an energy-transportation coordinated CMIES model, revealing the operational boundary without solving the NP-hard Minkowski sum problem.

\item {\color{black}A data-driven FRA method is proposed for CMIES aggregation, addressing imperfect information from unknown parameters and data privacy in coal mines.}
\end{enumerate}

 The reminder of the paper is organized as follows. Section II briefly introduces the coal mine VPP FRA problem. Section III formulates the inverse optimization model of coal mine VPP aggregation with the energy-transportation coordinated CMIES model. Section IV proposes the learning-based FRA method. Case studies are given in Section V with performance analysis. In the end, Section VI concludes this paper.

\section{Problem Description}
% In this section we briefly introduce the FRA problems with imperfect information in the coal mine VPP.
\subsection{Coal Mine VPP FRA}
{\color{black}We consider a typical scenario where the coal mine VPP aggregates several CMIESs operate in the distribution network, equipped with essential devices such as generation units, belt conveyors, and energy storage systems. This setup enables the CMIES to offer flexibility by managing its energy resources dynamically. Such flexibility is crucial for helping the distribution network integrate renewable energy and perform peak shaving and valley filling. To quantify this coal mine VPP FRA, we define it rigorously as follows.}
% The coal mine VPP aggregates flexible resources, mainly generation units and belt conveyors, to participate in the energy market and provide peak regulation. However, variable underground conditions lead to unknown deviations in belt conveyor parameters from their design values. Accurately approximating the feasible region of the coal mine VPP is crucial for effective market bidding, defined by its internal energy management as follows.
\begin{define}
  The feasible region of the coal mine VPP is a sub-space of the energy dispatching variables as the range of its power exchange with DSO and belt-conveyors load peak-valley regulation, noted as $\Omega_{\text{V}}$.
\end{define}
  \begin{align}\label{definition}
       \Omega_{\text{V}}:=\{(p_{{\text{BC}},|{\rm I}|}^{|{\rm T}|},p_{g}^{|{\rm T}|})|{\rm s.t.} &h(p_{{\text{BC}},|{\rm I}|}^{|{\rm T}|},p_{g}^{|{\rm T}|},x_{|{\rm I}|}^{|{\rm T}|},\Xi_{|{\rm I}|}) \hspace{-1mm}= \hspace{-1mm}0, \nonumber
      \\ &g(p_{{\text{BC}},|{\rm I}|}^{|{\rm T}|},p_{g}^{|{\rm T}|},x_{|{\rm I}|}^{|{\rm T}|},\Xi_{|{\rm I}|}) \hspace{-1mm}\leq \hspace{-1mm} 0 \}
  \end{align}
where $p_{{\text{BC}},|{\rm I}|}^{|{\rm T}|}$ denotes the total power consumption of BCs coal mine indexed by $i\in {\cal I}^{|{\rm I}|}$ at all time intervals $t\in {\cal T}^{|{\rm T}|}$. $p_{g}^{|{\rm T}|}$ is the total power exchange of the coal mine VPP with the DSO. $x_{|{\rm I}|}^{|{\rm T}|}$ and $\Xi_{|{\rm I}|}$ are the other decision variables and parameters in coal mine optimal energy dispatching. 

However, as pointed out, {\color{black}the parameters $\Xi_{|{\rm I}|}$ are usually unknown to the aggregator. Also, the vast number of equipment in the coal transportation network makes the of $\Xi_{|{\rm I}|}$ a high-dimension vector.} As defined in Eq. \eqref{definition}, the feasible region is a projection of the aggregated power exchange and peak-valley regulation capacity on the original feasible set, which is a NP-hard problem. Alternatively, one effective way is to leverage the historical optimal dispatching data to learn the surrogate feasible region, denoted as.
\begin{align}\label{surrogate}
     \hspace{-2mm}  {\tilde \Omega_{\text{V}}(\tilde \Xi_{|{\rm I}|})} \hspace{-1mm}:= \hspace{-1mm}\{(p_{{\text{BC}} \hspace{-1mm},|{\rm I}|}^{|{\rm T}|},p_{g}^{|{\rm T}|})|{\rm s.t.} & \tilde h(p_{{\text{BC}},|{\rm I}|}^{|{\rm T}|},p_{g}^{|{\rm T}|},\tilde x_{|{\rm I}|}^{|{\rm T}|},\tilde \Xi_{|{\rm I}|}) \hspace{-1mm}= \hspace{-1mm}0, \nonumber
      \\ &\tilde g(p_{{\text{BC}},|{\rm I}|}^{|{\rm T}|},p_{g}^{|{\rm T}|},\tilde x_{|{\rm I}|}^{|{\rm T}|},\tilde \Xi_{|{\rm I}|}) \hspace{-1mm}\leq \hspace{-1mm} 0 \}
  \end{align}
Therefore, the idea behind the surrogate model in Eq. \eqref{surrogate} is to use the approximated $\tilde \Xi_{|{\rm I}|}$ for getting a $\tilde \Omega_{\text{V}}$ close to the real feasible region $\Omega_{\text{V}}$ without knowing the true value of $\Xi_{|{\rm I}|}$. To this end, the coal mine VPP FRA can be formulated as.

\begin{align}\label{FRA}
 &\hspace{-14mm}\min_{\tilde \Xi_{|{\rm I}|}\in  [\underline\Xi_{|{\rm I}|}, \bar\Xi_{|{\rm I}|}]}{\mathcal L} \left( \tilde \Omega_{\text{V}}(\tilde \Xi_{|{\rm I}|}),\Omega_{\text{V}} \right)\\
 \hspace{8mm}{\text {s.t.}}\hspace{4mm}&h(p_{{\text{BC}},|{\rm I}|}^{|{\rm T}|},p_{g}^{|{\rm T}|},\tilde x_{|{\rm I}|}^{|{\rm T}|},\tilde \Xi_{|{\rm I}|}) \hspace{-1mm}= \hspace{-1mm}0 \nonumber \\
 &g(p_{{\text{BC}},|{\rm I}|}^{|{\rm T}|},p_{g}^{|{\rm T}|},\tilde x_{|{\rm I}|}^{|{\rm T}|},\tilde \Xi_{|{\rm I}|}) \hspace{-1mm}\leq \hspace{-1mm} 0 \nonumber
\end{align}
where ${\mathcal L}$ is a loss function to minimize the assessment error between $\tilde \Omega_{\text{V}}(\tilde \Xi_{|{\rm I}|})$ and $\Omega_{\text{V}}$. Hence, the feasible region can be approximated by solving the optimization problem Eq. \eqref{FRA} with historical data. The tractability of Eq. \eqref{FRA} relies on the original problem, which will be discussed in the following section.
 
\section{Energy-Transportation Coordinated Coal Mine Optimal Scheduling}
\subsection{CMIES Configuration}
The configuration of the CMIES and the coal transportation network (CTN) is illustrated in Fig. 1. This coordinated energy-transportation system includes wind turbines (WTs), photovoltaic systems (PVs), combined heat and power units (CHPs), microturbines (MTs), regenerative thermal oxidizers (RTOs), and water source heat pumps (WSHPs). Energy storage is provided through the pumped-hydro storage (PHS) and thermal storage tanks (TSTs). These components are integrated to supply both electrical and thermal energy. To facilitate demand response in coal transportation, the CTN is equipped with belt conveyors (BCs) at various levels and silos, utilizing electricity from the CMIES to transport raw coal from the coal face to the coal preparation plant (CPP) according to CTN scheduling.

\begin{figure}[!t]
\centering
\includegraphics[width=0.95\linewidth]{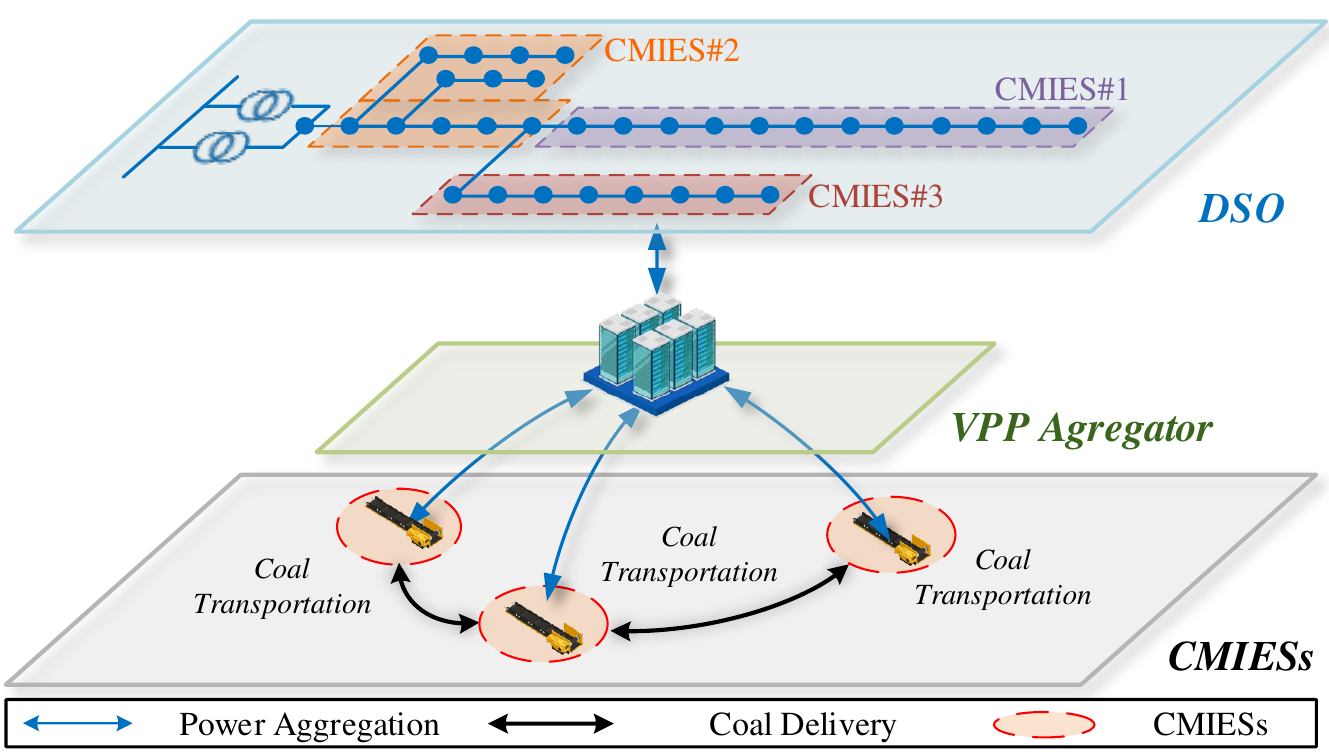}\vspace{-2mm}
\caption{Coal Mine VPP Aggregation with CMIESs}\vspace{-6mm}
\label{fig_1}
\end{figure}

\subsection{Objective Function}
The objective function aims to minimize the operational costs of the coal mine, incorporating expenses from energy trading under the time-varing price $\kappa_{\text{Price}}$, fuel costs for generation, and maintenance expenses.
\begin{subequations}
\begin{align}
    C_{\text{Sys}} &= \sum_{t \in {\cal T}} \left( \kappa_{\text{Price}} p_{\text{grid}}^{t} + \kappa_{\text{CG}}^{fu} p_{\text{CHP}}^{t}+C_{\text{OM}}^{t}\right)\\
    C_{\text{OM}}^{t} &= \kappa_{\text{PV}}^{om} p_{\text{PV}}^{t} + \kappa_{\text{WT}}^{om} p_{\text{WT}}^{t} +  \kappa_{\text{GT}}^{om,e} p_{\text{GT}}^{t,e} +\kappa_{\text{CHP}}^{om} p_{\text{CHP}}^{t,f} \notag \\
    & \kappa_{\text{RTO}}^{om} p_{\text{RTO}}^{t} + \sum_{j \in N_{\text{BC}}} \kappa_{\text{BC}}^{om,h} p_{\text{BC}}^{t,h} + \kappa_{\text{PHS}}^{om,h} \left( p_{\text{PHSC}}^{t} + p_{\text{PHSD}}^{t} \right) \notag \\
    & \quad + \kappa_{\text{TST}}^{om,h} \left( H_{\text{TSC}}^{t} + H_{\text{TSD}}^{t} \right) + \kappa_{\text{WSHP}}^{om,h} p_{\text{WSHP}}^{t}, \quad \forall t \in {\cal T}
\end{align}
\end{subequations}
\subsection{Constraints}
\subsubsection{Coal Transportation}
The CTN consists of coal faces, BCs, silos, and the CPP, arranged in a radial topology. The CTN can be modeled as a graph ${\ G}_{\text {CTN}}=\{{\sigma_{\text {L}}},T_{\text {N}} \}$, where $\sigma_{\text {L}}$ and $T_{\text {N}}$ represent the coal transportation links and nodes, respectively. The node set $T_{\text {N}}$ comprises multiple hierarchical nodes, including the coal face supply node $T_{\text {CF}}$, coal silo storage node $T_{\text {SS}}$, transfer node $T_{\text {MS}}$, and the CPP demand node $T_{\text {CPP}}$, expressed as $T_{\text {N}}=\{T_{\text {CF}}\cup T_{\text {SS}}\cup T_{\text {MS}}\cup T_{\text {CPP}} \}$. Each BC branch supports coal mass transfer as a flow between nodes at different levels, represented as $\sigma_{\text {L}}=\{\sigma_{\text {CF,SS}}\cup\sigma_{\text {SS,MS}}\cup \sigma_{\text {MS,CPP}}\}$. Coal mass flow proceeds1 from the coal face to the CPP. The CTN model is formulated as follows.
\begin{subequations}
\begin{align}
{\sigma}_{l,m}^{t}=\frac{{\cal Q}_{l,m}^{t}}{3.6*\text{V}_{\text{BC}}},&\forall t \in {\cal T},\forall {\sigma}_{l,m} \in {\sigma}_{\text {CF,SS}}\label{coal_face}\\
\underline{\sigma}_{l,m}^{ramp}\leq {\sigma}_{l,m}^{t}-{\sigma}_{l,m}^{t-1} \leq \bar{\sigma}_{l,m}^{ramp},&\forall t \in {\cal T},\forall {\sigma}_{l,m} \in {\sigma}_{\text {CF,SS}}\label{feedrate_ramp_minmax}\\
\sum_{l}{\sigma}_{l,m}^{t} \leq {\sigma}_{m,n}^{t}\leq \bar {\sigma}_{m,n},&\forall t \in {\cal T},\forall {\sigma}_{l,m} \in {\sigma}_{\text {CF,SS}},\nonumber \\  &\forall {\sigma}_{l,m} \in {\sigma}_{\text {SS,MS}}\label{feedrate_minmax}\\
\sum_{t}{\cal Q}_{l,m}^{t}\leq \sum_{t}{\cal Q}_{n,p}^{t} \leq {\cal Q}_{\text{L,CPP}},&\forall t \in {\cal T},\forall {\sigma}_{l,m} \in {\sigma}_{\text {CF,SS}},\nonumber \\
&\forall {\sigma}_{n,p} \in {\sigma}_{\text {MS,CPP}}\label{feedrate_balance}
\end{align}
\end{subequations}
where ${\cal Q}_{l,m}^{t}$ and ${\sigma}_{l,m}^{t}$ represent the feed rate and coal mass in coal transportation between coal faces and shaft silos respectively. $\underline{\sigma}_{l,m}^{ramp}$ and $\bar{\sigma}_{l,m}^{ramp}$ represent the minimal and maximal ramp limits of coal mass delivery respectively. $\bar {\sigma}_{m,n}$ denotes the maximal limits of coal mass delivered from shaft silos to the main silo. ${\cal Q}_{\text{L,CPP}}$ is the coal mass load in the CPP. Since coal mass flows only in one direction, $l, m, n,$ and $p$ represent the CTN nodes $T_{\text {CF}}, T_{\text {SS}}, T_{\text {MS}}$ and $T_{\text {CPP}} $ with the sequence from the coal face to the CPP. 
% Eq. \eqref{coal_face} denotes the relationship between the unit coal flow mass and feed rate. Eq. \eqref{feedrate_ramp_minmax} ensures that the ramp-up and down of coal mass flow on each link should vary within a limited range. For coal transportation safety, Eq. \eqref{feedrate_minmax} shows the coal delivery safety rule for preventing the BCs from piling raw coal in each transportation junction part. Eq. \eqref{feedrate_balance} enforces that the delivered raw coal from the coal face can meet the demand of the central CPP. Typically, the proposed model can be generally applied to any coal mines where the BCs are the main equipment for coal delivery.

To describe the virtual energy storage characteristic in the CTN, the model of a coal silo is equivalent to the battery model without considering decay. The model is described as:
\begin{subequations}
\begin{align}\label{coal_silo}
 M_{{\text {Silo,}k}}^{t+1}=&M_{{\text {Silo,}k}}^{t}+\sigma_{\text {BC,}jk}^{t}-\sigma_{\text {BC,}kj}^{t},\forall t \in {\cal T},\forall k \in {\text N}_{{\text {Silo}}} \\
 &\underline M_{{\text {Silo}}}\leq M_{{\text {Silo,}k}}^{t} \leq \bar M_{{\text {Silo}}}, \hspace{1.75mm}\forall t \in {\cal T},\forall k \in {\text N}_{{\text {Silo}}} \label{coal_silo_min_max}\\
 &M_{{\text {Silo}}}^{{\text {1}}}=M_{{\text {Silo}}}^{\text {st}}, \hspace{4mm}M_{{\text {Silo}}}^{{\text {24}}}=M_{{\text {Silo}}}^{\text {end}}
\end{align}
\end{subequations}

\subsubsection{Energy Units}
For the electrical and thermal power coupled energy units, the RTO, the CHP, the GT and the WSHP model are formulated as follows.
\begin{subequations}
\begin{align}
&p_{X}^{t}={\text{EHR}}_{X}*h_{X}^{t},\forall t \in {\cal T} \\
&\underline h_{{X}}<h_{X}^{t}<=\bar h_{X},\forall t \in {\cal T}\\
&X \in \{{\text{RTO,CHP,GT,WSHP}}\}
\end{align}
\end{subequations}
where ${\text{EHR}}_{X}$ is the electricity and heat generation ratio of unit $X$. 
\subsubsection{Belt Conveyors}
BCs carry produced raw coal from the work face to the CPP by consuming electricity. The electric power consumption of BCs can be represented using a generalized coal transportation model based on well-known standards or specifications, such as ISO 5048, DIN 22101, JIS B 8805, formulated as follows.
\begin{subequations}
 \begin{align}\label{BC_Power}
   p_{{{\text {BC}}},j}^{t} &= cof_{{\text {BC}}} \left[ \theta_{2,j}{\text {V}_{\text {BC}}} + \left(\theta_{4,j} +  \frac{\text{V}_{\text{BC}}}{\text{3.6}}\right){\cal Q}_{{{\text {BC}}},j}^{t} \right], \nonumber \\
   &\hspace{4cm}\forall t \in {\cal T}, \forall j \in {{\text {N}}_{\text {BC}}}
\end{align}
\end{subequations}

\subsubsection{Energy Storage Units}
The PHS systems established in abandoned coal mine goaves provide flexible options for storing and releasing electrical energy. Similar to PHS, TSTs are also utilized to mitigate peak thermal loads. The energy storage units are represented by Eqs. \eqref{PHS} and \eqref{TST}.
\begin{subequations}\label{PHS}
\begin{align}
   &E_{{\text{PHS}}}^{t+1}=\gamma_{{\text{PHS}}}E_{{\text{PHS}}}^{t}+\eta_{{\text{PHS}}}\left(p_{{\text{PHSC}}}^{t}-p_{{\text{PHSD}}}^{t}\right),\forall t \in {\cal T}\\
&E_{{\text{PHS}}}^{t}\in [\underline E_{{\text{PHS}}},\bar E_{{\text{PHS}}}],\hspace{0mm}\\
&p_{{\text{PHSC}}}^{t}\in [\underline p_{{\text{PHSC}}},\bar p_{{\text{PHSC}}}],p_{{\text{PHSD}}}\in [\underline p_{{\text{PHSD}}},\bar p_{{\text{PHSD}}}],\hspace{2mm}\\
&E_{{\text{PHS}}}^{{\text{1}}}=E_{{\text{PHS}}}^{{\text{st}}}, \hspace{4mm}E_{{\text{PHS}}}^{{\text{24}}}=E_{{\text{PHS}}}^{{\text{end}}}
\end{align}
\vspace{-6mm}
\end{subequations}
\begin{subequations}\label{TST}
\begin{align}
   &E_{{\text{TST}}}^{t+1}=\gamma_{{\text{TST}}}E_{{\text{TST}}}^{t}+\eta_{{\text{TST}}}\left(h_{{\text{TSTC}}}^{t}-h_{{\text{TSTD}}}^{t}\right),\forall t \in {\cal T}\\
&E_{{\text{TST}}}^{t}\in [\underline E_{{\text{TST}}},\bar E_{{\text{TST}}}],\hspace{0mm}\\
&h_{{\text{TSTC}}}^{t}\in [\underline h_{{\text{TSTC}}},\bar h_{{\text{TSTC}}}],h_{{\text{TSTD}}}\in [\underline h_{{\text{TSTD}}},\bar h_{{\text{TSTD}}}],\hspace{2mm}\\
&E_{{\text{TST}}}^{{\text{1}}}=E_{{\text{TST}}}^{{\text{st}}}, \hspace{4mm}E_{{\text{TST}}}^{{\text{24}}}=E_{{\text{TST}}}^{{\text{end}}}
\end{align}
\end{subequations}

\subsubsection{Energy Balance}
The electrical and thermal power balance constraints for the coal mines are formulated as follows.
\begin{subequations}
\begin{align}\label{power_balance}
    &\underline p_{g} \leq p_{g}^{t} \leq \bar p_{g}, \forall t \in {\cal {T}}\\
    &p_{g}^{t} + p_{\text{RTO}}^{t} + p_{\text{CHP}}^{t} + p_{\text{GT}}^{t} + p_{\text{PV}}^{t} + p_{\text{WT}}^{t} + p_{\text{PHSD}}^{t} \notag \\
    & \hspace{8.25mm}= \; p_{\text{PHSC}}^{t} + p_{\text{Load}}^{t} + \sum_{j=1}^{N_{\text{BC}}} p_{\text{BC},j}^{t} + p_{\text{WSHP}}^{t}, \forall t \in {\cal {T}}\\
    &\sum_{X} h_{\text {X}}^{t}+h_{\text {TSTD}}^{t}=h_{\text {TSTC}}^{t}+h_{\text {Load}}^{t}, \forall t \in {\cal T}\label{heat_balance}
\end{align}
\end{subequations}
\subsection{Model Reformulation}
In the above model, the parameters are $\Xi_{|{\rm I}|}=\left[ \underline p_{\text{BC},j}, \bar p_{\text{BC},j}, \underline p_{g}, \bar p_{g}, \theta_{\text{2,j}}, \forall j \in {{\text {N}}_{\text {BC}}} \right]$ and $x_{|{\rm I}|}^{|{\rm T}|}$ are other variables except $p_{{\text{BC}},|{\rm I}|}^{|{\rm T}|}$ and $p_{g}^{|{\rm T}|}$. The model is reformulated into an impact form in Eq. \eqref{Reformulation}.
\begin{align}\label{Reformulation}
 &\hspace{-4mm}\min_{p_{{\text{BC}},|{\rm I}|}^{|{\rm T}|},p_{g}^{|{\rm T}|},x_{|{\rm I}|}^{|{\rm T}|}}C_{\text{Sys}}\\
 &{\hspace{4mm}{\text {s.t.}}\hspace{4mm}{\text {Eqs.~}} \eqref{coal_face} \sim \eqref{heat_balance}}\nonumber
 \end{align}

\section{Solution Methodology}
\subsection{\color{black}Parameter Identification}
Based on historical data set $D=\left[p_{\text{BC}, |I|}^{{\text {D}}},p_{\text{g}}^{{\text {D}},|T|}\kappa_{\text{Price}}^{\text {D}} \right]$, the bilevel optimization problem \eqref{FRA} can be converted into a single level problem via KKT conditions. Thus, the data-driven inverse optimization based FRA can be formulated as follows.
\begin{subequations}\label{FRA_DDIO}
\begin{align}
 &\hspace{-16mm}\min_{\tilde \Xi_{|{\rm I}|}\in  [\underline\Xi_{|{\rm I}|}, \bar\Xi_{|{\rm I}|}]}{\mathcal L}=\left\| p_{{\text{BC}}}-p_{\text{BC}, |I|}^{{\text {D}},|T|}\right\|_{2}+\left\|p_{{\text{g}}}-p_{\text{g}}^{{\text {D}},|T|}\right\|_{2}\\
 \hspace{8mm}{\text {s.t.}}\hspace{4mm}&\frac{\partial \mathcal{L}}{\partial p_{\text{BC}, |I|}^{|T|}} = 0, \quad \frac{\partial \mathcal{L}}{\partial p_g^{|T|}} = 0, \quad \frac{\partial \mathcal{L}}{\partial x_{|I|}^{|T|}} = 0 \label{FRA_Stationarity}\\
    & h(p_{\text{BC}, |I|}^{|T|}, p_g^{|T|}, x_{|I|}^{|T|}, \Xi_{|I|}) = 0, \label{FRA_Primal_Eq}\\
    & g(p_{\text{BC}, |I|}^{|T|}, p_g^{|T|}, x_{|I|}^{|T|}, \Xi_{|I|}) \leq 0,\label{FRA_Primal_Ineq}\\
    & \mu \geq 0, \mu^T  \perp g(p_{\text{BC}, |I|}^{|T|}, p_g^{|T|}, x_{|I|}^{|T|}, \Xi_{|I|}),\label{FRA_Dual_Com} \\   
    & \mathcal{L}^*(\lambda,\mu) = \mathcal{L}\label{FRA_Strong_Dual}
\end{align}
\end{subequations}
where Eqs. \eqref{FRA_Stationarity}-\eqref{FRA_Strong_Dual} are the stationarity, primal feasiblility, dual feasibility, complementary slackness and strong duality conditions respectively. By solving this problem with finite number of data in $D$, the $\Omega_{\text{V}}$ can be effectively approximated.

\subsection{ Learning-based FRA algorithm}
Although the nonlinear complementary slackness condition in \eqref{FRA_Dual_Com} can be handled with the Big M method, it still contains a large number of binary variables. This would leads to a heavy computation burden especially when the original CMIES dispatching problem has high-dimension variables and numerous constraints. One effective approach is to leverage the learning method to get the FRA solution. 
To solve the nonlinear FRA problem in Eq. \eqref{FRA_DDIO}, the LFRA algorithm is proposed to solve the following inverse optimization problem \eqref{LFRA} with the dynamic updated data. {\color{black}Detailed procedures are given in Algorithm 1. }

\begin{subequations}\label{LFRA}
\begin{align}
 &\hspace{-16mm}\min_{\tilde \Xi_{|{\rm I}|}^{\xi}}\left({\mathcal L}+\frac{\rho}{2}\left\|\tilde \Xi_{|{\rm I}|}^{\xi}-\tilde\Xi_{|{\rm I}|}^{\xi-1}\right\|_{2}\right), \forall \xi\\
 \hspace{8mm}{\text {s.t.}}\hspace{4mm}&{\text {Eqs.~}} \eqref{FRA_Stationarity} \sim \eqref{FRA_Strong_Dual}
\end{align}
\end{subequations}

\begin{algorithm}
\SetKwInOut{In}{Input}
\SetKwInOut{Out}{Output}
\In{\text{Iteration index} $\xi=1$, \text{Penalty factor} $\rho$, \text{Tolerance value} $\epsilon$, \text{Historical data}  $D$ }
\text{\textit{Initialize:} Solve the FRA in \eqref{FRA_DDIO} with parameter $\tilde \Xi_{|{\rm I}|}^{0}$}

\While{$ \left\|{\mathcal L}^{\xi}-{\mathcal L}^{\xi-1} \right\|_{2}\geq \epsilon  $}{
 \For{$s=\xi$ \KwTo $\xi+|D|$ }  {Solve \eqref{LFRA} to obtain $\tilde \Xi_{|{\rm I}|}^{\xi,s}$} 
 Update the $\tilde \Xi_{|{\rm I}|}^{\xi}=\frac{1}{|D|}\sum\limits_{s=\xi}^{\xi+|D|}\tilde \Xi_{|{\rm I}|}^{\xi,s} $

    $\xi=\xi+1$ 
  }        
\Out{$\tilde \Omega_{\text{V}}(\tilde \Xi_{|{\rm I}|})$}
\caption{Learning-based FRA Algorithm} \label{algo_market}
\end{algorithm} \vspace{-2mm}
The LFRA algorithm iteratively optimizes parameters for high-dimensional systems. It processes scenario-specific data in a mini-batch style, updating parameters and averaging results for robustness. Convergence is achieved when the loss change meets a tolerance. This incremental learning based approach enhances adaptability and convergence efficiency.

\section{Case Studies}
\vspace{-0mm}
\subsection{Test System Setup}
The proposed LFRA method is numerically evaluated using the data of coal mines in IEEE 33-bus distribution system, illustrated in the Fig. 2. The FRA is solved on the MATLAB R2023b platform via Gurobi 11.0.0 using an Apple Silicon M1 CPU with 16GB RAM.
\begin{figure}
\centering
\includegraphics[width=0.95\linewidth]{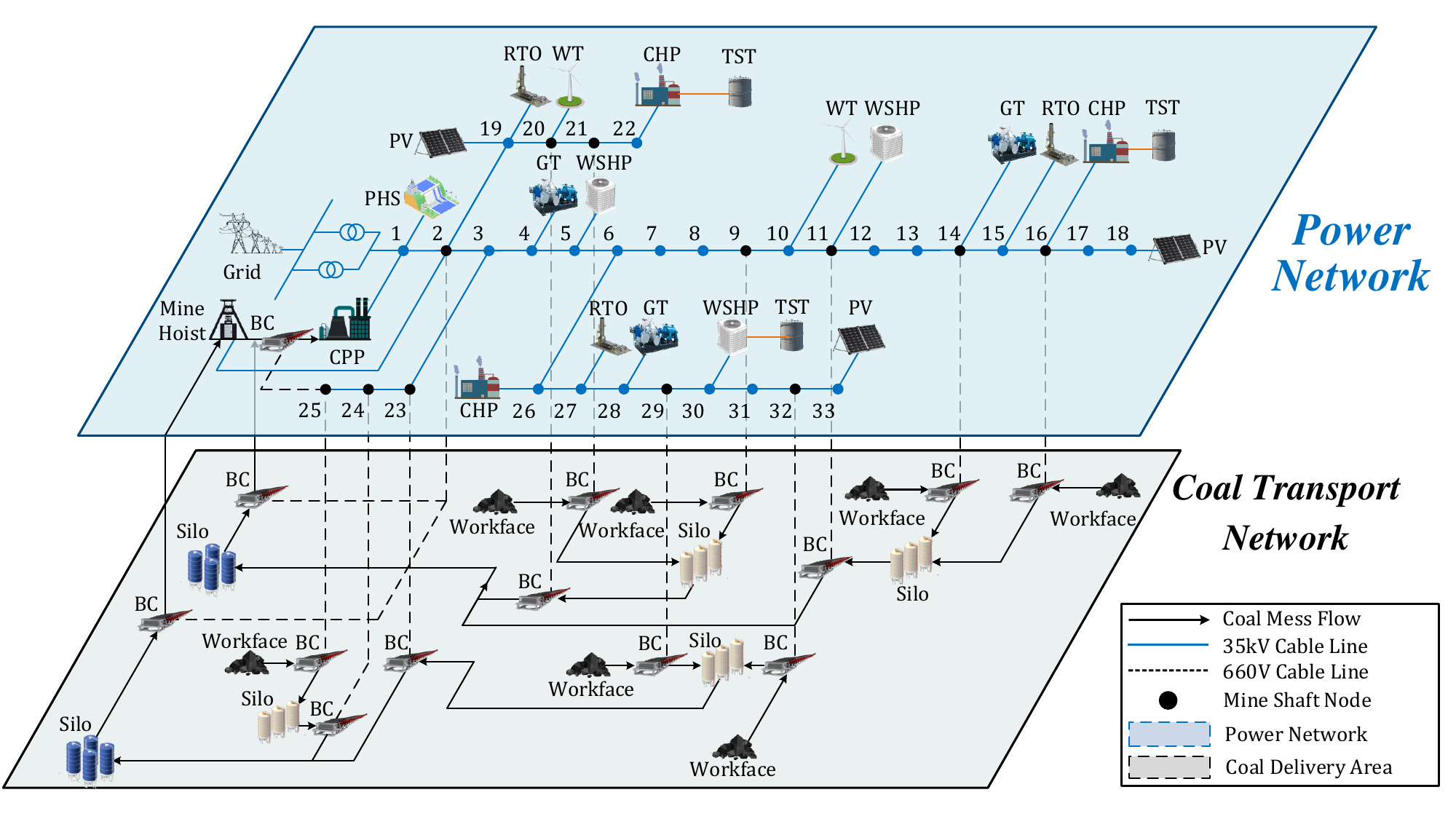}\vspace{-2mm}
\caption{System configuration}\vspace{-2mm}
\label{fig_2}
\end{figure}
\subsection{Parameter Identification}
 Fig. 3 illustrates the parameter $\theta_2$ identification results and percentage relative errors for individual BC devices over training, revealing that the proposed LFRA can effectively approximate the unknown parameters from historical data. It is represented that there are noticeable fluctuations at the beginning of the training, which gradually reduce, indicating that the model is learning the parameters to better fit the actual ones. Near the end of training, most approximated $\theta_2$ for diverse BCs maintain low error rates, which are under 1\% and relatively acceptable in application.
\begin{figure}
\centering
\includegraphics[width=0.95\linewidth]{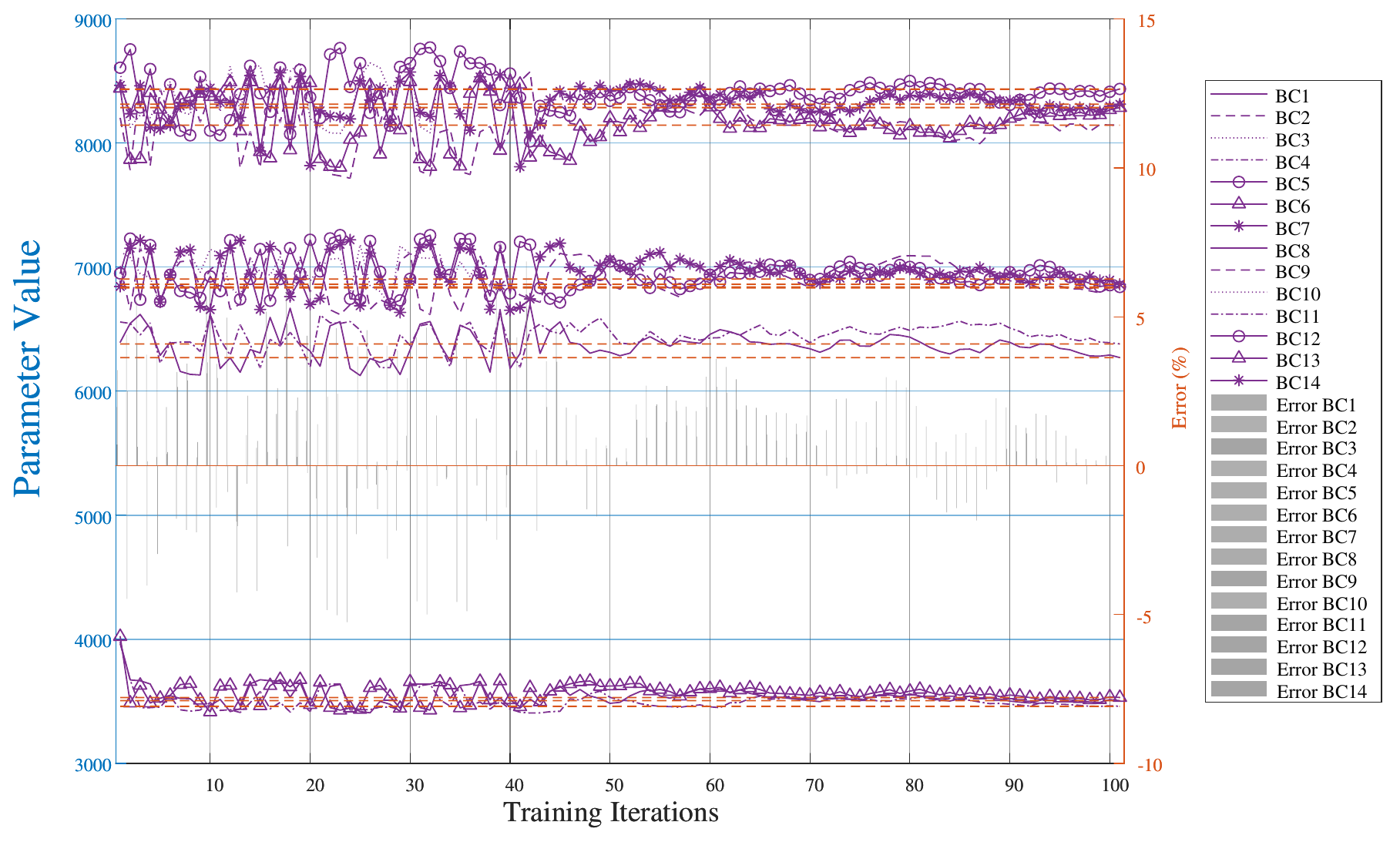}\vspace{-0mm}
\caption{The parameter identification results of 14 BCs}\vspace{-0mm}
\label{fig_4}
\end{figure}
\subsection{Performance Comparison}
The maximum and minimum power values for BCs and power exchange within the VPP, predicted by the proposed method, are compared with {\color{black}true values} in Fig. 4. 

Results demonstrate that the proposed LFRA method effectively approximates both the maximum and minimum power bounds for belt conveyors and power exchange, validating its accuracy and reliability. The low error observed in the belt conveyor limits indicates that the method can precisely capture the stable operational characteristics of these components, likely due to the predictable nature of their power profiles. 
For belt conveyors power approximation, result shows an a final error below 1\%, while the minimum power limit maintains an error of under 0.3\% across all data points. In the case of power exchange, the approximation achieves an average error of approximately 3\% for both maximum and minimum power bounds. This indicates that the proposed method can ensure the accuracy of FRA with historical data.
\begin{table*}[]
\centering
\caption{Result Comparison on Approximation Error}
\begin{tabular}{c|cc|cc|cc|cc|cc}
\hline
\multirow{2}{*}{Method} &
  \multicolumn{2}{c|}{\(\bar p_{\text{BC}}\)} &
  \multicolumn{2}{c|}{\(\underline p_{\text{BC}}\)} &
  \multicolumn{2}{c|}{\(\bar p_{g}\)} &
  \multicolumn{2}{c|}{\(\underline p_{g}\)} &
  \multicolumn{2}{c}{\(\theta_2\)} \\ \cline{2-11} 
 &
  \multicolumn{1}{c|}{LSTM} &
  LFRA &
  \multicolumn{1}{c|}{LSTM} &
  LFRA &
  \multicolumn{1}{c|}{LSTM} &
  LFRA &
  \multicolumn{1}{c|}{LSTM} &
  LFRA &
  \multicolumn{1}{c|}{LSTM} &
  LFRA \\ \hline
RMSE (\%) &
  \multicolumn{1}{c|}{25.34} &
  2.95 &
  \multicolumn{1}{c|}{NaN} &
  NaN &
  \multicolumn{1}{c|}{7.32} &
  3.06 &
  \multicolumn{1}{c|}{10.8} &
  0.11 &
  \multicolumn{1}{c|}{7.39} &
  2.21 \\ \hline
MAE (\%) &
  \multicolumn{1}{c|}{25.02} &
  5.28 &
  \multicolumn{1}{c|}{NaN} &
  NaN &
  \multicolumn{1}{c|}{7.32} &
  5.20 &
  \multicolumn{1}{c|}{10.8} &
  0.13 &
  \multicolumn{1}{c|}{7.13} &
  1.71 \\ \hline
\end{tabular}
\end{table*}

\begin{figure}[!t]
\centering
\includegraphics[width=1.0\linewidth]{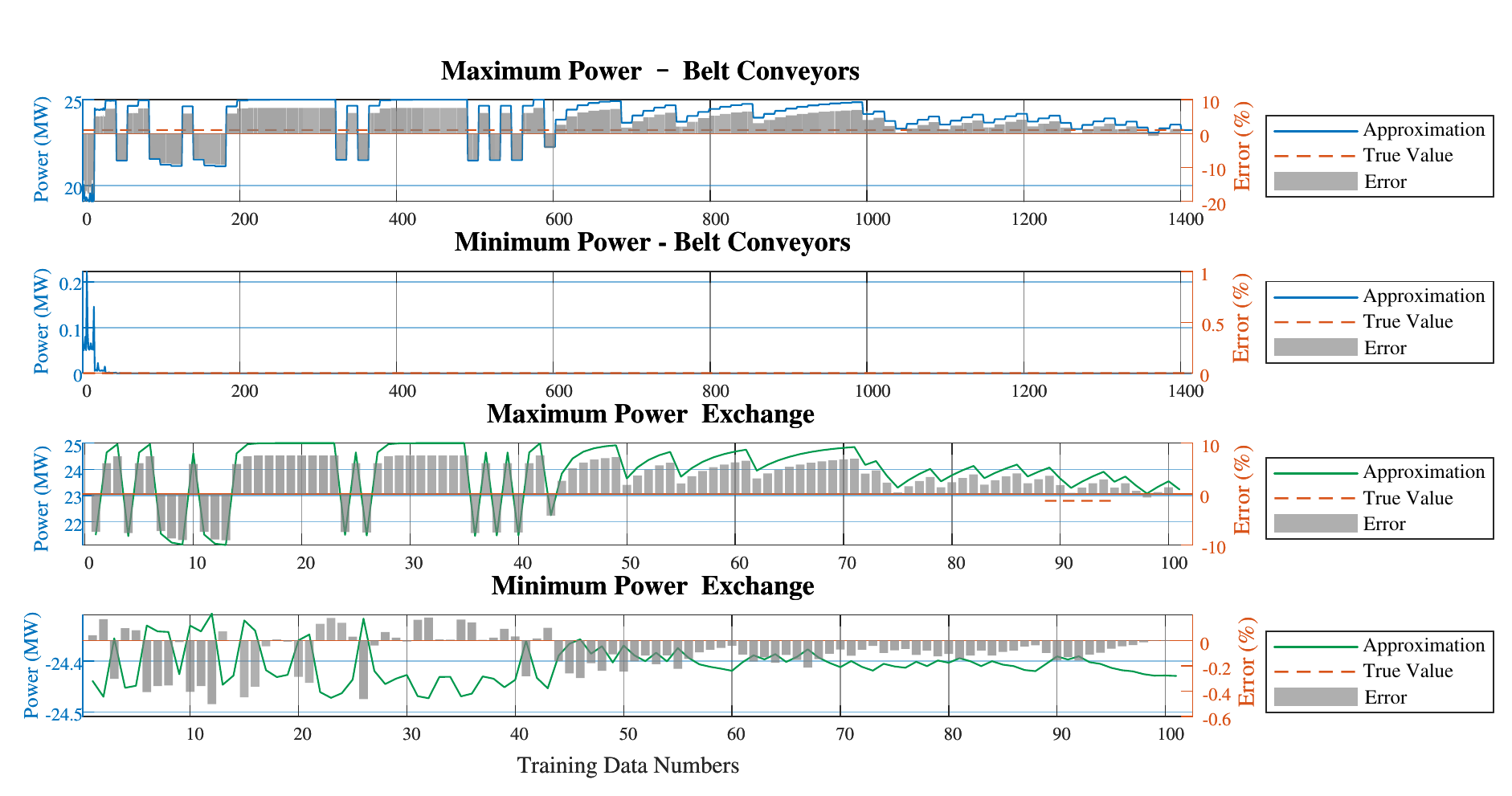}\vspace{-2mm}
\caption{The FRA and error results }\vspace{-2mm}
\label{fig_3}
\end{figure}

Table I presents a comparative analysis of approximation errors using two methods: LSTM and the proposed LFRA. The results show that the LFRA consistently outperforms LSTM across all evaluated parameters, achieving significantly lower RMSE and MAE values. Specifically, for belt conveyor maximum power , LFRA yields an RMSE of 2.95\% and MAE of 5.28\%, markedly lower than LSTM’s RMSE and MAE values of 25.34\% and 25.02\%, respectively. This indicates LFRA’s superior accuracy in capturing the peak-valley regulation potential of the coal mine VPP. However, due to the minimum BC power can reach 0, the RMSE and MAE for both LSTM and LFRA are NaN. Notably, when approximating $\bar p_{g}$ and $\underline p_{g}$,LFRA maintains accurate approximations, while LSTM shows higher RMSE and MAE. For the parameter $\theta_2$, LFRA achieves 2.21\% RMSE and 1.71\% MAE, while LSTM has higher errors, further confirming LFRA’s accuracy advantage. These results validate LFRA’s effectiveness for high-accuracy FRA in the coal mine VPP.

\section{Conclusion}
In this study, a learning-based FRA method was developed for the coal mine VPP under imperfect information. The proposed LFRA approach demonstrates superior accuracy in approximating operational limits compared to traditional methods, achieving significantly lower RMSE and MAE across key parameters. Future work will focus on extending LFRA for real-time applications and addressing nonlinear characteristics in industrial FRA.

% \begin{enumerate}

% \item Our model could effectively optimize investment costs while simultaneously enhancing system resilience.

% \item The proposed method effectively handles non-anticipativity constraints by developing pre-event plans that deploy additional equipment to improve resilience. 

% \item The proposed method demonstrates strong computational efficiency, as shown by reduced solution times and fewer iterations.
% \end{enumerate}

\bibliographystyle{IEEEtran}
\bibliography{Mybib}

\begin{thebibliography}{10}
\providecommand{\url}[1]{#1}
\csname url@samestyle\endcsname
\providecommand{\newblock}{\relax}
\providecommand{\bibinfo}[2]{#2}
\providecommand{\BIBentrySTDinterwordspacing}{\spaceskip=0pt\relax}
\providecommand{\BIBentryALTinterwordstretchfactor}{4}
\providecommand{\BIBentryALTinterwordspacing}{\spaceskip=\fontdimen2\font plus
\BIBentryALTinterwordstretchfactor\fontdimen3\font minus \fontdimen4\font\relax}
\providecommand{\BIBforeignlanguage}[2]{{%
\expandafter\ifx\csname l@#1\endcsname\relax
\typeout{** WARNING: IEEEtran.bst: No hyphenation pattern has been}%
\typeout{** loaded for the language `#1'. Using the pattern for}%
\typeout{** the default language instead.}%
\else
\language=\csname l@#1\endcsname
\fi
#2}}
\providecommand{\BIBdecl}{\relax}
\BIBdecl

\bibitem{coalconsumption}
\BIBentryALTinterwordspacing
{National Bureau of Statics}, \emph{Stastical year book of China}.\hskip 1em plus 0.5em minus 0.4em\relax Beijing: China statics Press, 2020. [Online]. Available: \url{https://www.stats.gov.cn/sj/ndsj/2024/indexch.htm}
\BIBentrySTDinterwordspacing

\bibitem{lyu2023lstn}
R.~Lyu, H.~Guo, Y.~Zheng, Y.~Bai, and Q.~Chen, ``Lstn: a linear model of industrial production process for demand response,'' in \emph{2023 IEEE PES Innovative Smart Grid Technologies Europe (ISGT EUROPE)}.\hskip 1em plus 0.5em minus 0.4em\relax IEEE, 2023, pp. 1--5.

\bibitem{Tan_VB}
Z.~Tan, A.~Yu, H.~Zhong, X.~Zhang, Q.~Xia, and C.~Kang, ``Optimal virtual battery model for aggregating storage-like resources with network constraints,'' \emph{CSEE Journal of Power and Energy Systems}, 2022.

\bibitem{Industrial_VPP1}
H.~Zhao, B.~Wang, Z.~Pan, H.~Sun, Q.~Guo, and Y.~Xue, ``Aggregating additional flexibility from quick-start devices for multi-energy virtual power plants,'' \emph{IEEE Transactions on Sustainable Energy}, vol.~12, no.~1, pp. 646--658, 2021.

\bibitem{HUANG}
H.~Huang, R.~Liang, C.~Lv, M.~Lu, D.~Gong, and S.~Yin, ``Two-stage robust stochastic scheduling for energy recovery in coal mine integrated energy system,'' \emph{Applied Energy}, vol. 290, p. 116759, 2021.

\bibitem{VES}
H.~Huang, Z.~Li, H.~B. Gooi, H.~Qiu, X.~Zhang, C.~Lv, R.~Liang, and D.~Gong, ``Distributionally robust energy-transportation coordination in coal mine integrated energy systems,'' \emph{Applied Energy}, vol. 333, p. 120577, 2023.

\bibitem{MaJun}
J.~Ma, Y.~Zhang, Y.~Wang, D.~Gong, X.~Sun, and B.~Zeng, ``A multitask multiobjective operation optimization method for coal mine integrated energy system,'' \emph{IEEE Transactions on Industrial Informatics}, vol.~20, no.~9, pp. 11\,149--11\,160, 2024.

\bibitem{Blockchain_coalmine}
H.~Huang, Z.~Li, L.~P. M.~I. Sampath, J.~Yang, H.~D. Nguyen, H.~B. Gooi, R.~Liang, and D.~Gong, ``Blockchain-enabled carbon and energy trading for network-constrained coal mines with uncertainties,'' \emph{IEEE Transactions on Sustainable Energy}, vol.~14, no.~3, pp. 1634--1647, 2023.

\bibitem{JYP}
Y.~Jiang, Z.~Ren, and W.~Li, ``Committed carbon emission operation region for integrated energy systems: Concepts and analyses,'' \emph{IEEE Transactions on Sustainable Energy}, vol.~15, no.~2, pp. 1194--1209, 2024.

\bibitem{MKS}
Y.~Wen, Z.~Hu, S.~You, and X.~Duan, ``Aggregate feasible region of ders: Exact formulation and approximate models,'' \emph{IEEE Transactions on Smart Grid}, vol.~13, no.~6, pp. 4405--4423, 2022.

\bibitem{Pareek}
P.~Pareek, A.~Singh, L.~P. Mohasha, I.~Sampath, H.~B. Gooi, and H.~D. Nguyen, ``Privacy-preserving feasibility assessment for p2p energy trading and storage integration,'' in \emph{2022 IEEE Power \& Energy Society General Meeting (PESGM)}, 2022, pp. 1--5.

\bibitem{IO_1}
T.~C. Chan and N.~Kaw, ``Inverse optimization for the recovery of constraint parameters,'' \emph{European Journal of Operational Research}, vol. 282, no.~2, pp. 415--427, 2020.

\bibitem{Tan_DDIO}
Z.~Tan, Z.~Yan, Q.~Xia, and Y.~Wang, ``Data-driven inverse optimization for modeling intertemporally responsive loads,'' \emph{IEEE Transactions on Smart Grid}, vol.~14, no.~5, pp. 4129--4132, 2023.

\bibitem{Ruike}
R.~Lyu, H.~Guo, and Q.~Chen, ``Approximating energy-regulation feasible region of virtual power plants: a data-driven inverse optimization approach,'' in \emph{IEEE PES General Meeting}, 2024.

\end{thebibliography}
\end{document}